\documentclass[10pt,a4paper,english]{IEEEtran}
\usepackage{tgtermes}
\usepackage{helvet}

\usepackage[T1]{fontenc}
\usepackage[latin9]{inputenc}
\usepackage{amsbsy}
\usepackage{amssymb}
\usepackage{graphicx}

\makeatletter

\pdfpageheight\paperheight
\pdfpagewidth\paperwidth

\providecommand{\tabularnewline}{\\}

\makeatother

\usepackage{babel}
\begin{document}
\title{Weighted Particle-Based Optimization for Efficient Generalized Posterior
Calibration}
\author{Masahiro Tanaka\thanks{Faculty of Economics, Fukuoka University, Fukuoka, Japan. Address:
8-19-1, Nanakuma, Jonan, Fukuoka, Japan 814-0180. E-mail: gspddlnit45@toki.waseda.jp.
This study was supported by JSPS KAKENHI Grant Number 20K22096.}}
\maketitle
\begin{abstract}
In the realm of statistical learning, the increasing volume of accessible
data and increasing model complexity necessitate robust methodologies.
This paper explores two branches of robust Bayesian methods in response
to this trend. The first is generalized Bayesian inference, which
introduces a learning rate parameter to enhance robustness against
model misspecifications. The second is Gibbs posterior inference,
which formulates inferential problems using generic loss functions
rather than probabilistic models. In such approaches, it is necessary
to calibrate the spread of the posterior distribution by selecting
a learning rate parameter. The study aims to enhance the generalized
posterior calibration (GPC) algorithm proposed by \cite{Syring2019}.
Their algorithm chooses the learning rate to achieve the nominal frequentist
coverage probability, but it is computationally intensive because
it requires repeated posterior simulations for bootstrap samples.
We propose a more efficient version of the GPC inspired by sequential
Monte Carlo (SMC) samplers. A target distribution with a different
learning rate is evaluated without posterior simulation as in the
reweighting step in SMC sampling. Thus, the proposed algorithm can
reach the desirable value within a few iterations. This improvement
substantially reduces the computational cost of the GPC. Its efficacy
is demonstrated through synthetic and real data applications.
\end{abstract}

\section{Introduction}

As the volume of accessible data and the complexity of models increase,
the necessity for robustness in statistical learning intensifies.
With this trend in mind, this study delves into two branches of robust
Bayesian methods. First, it focuses on generalized Bayesian inference
\cite{Martin2022,Nottforthcoming}. Let $\mathcal{D}$ represent a
dataset comprising $N$ independent samples, denoted by $\mathcal{D}=\left\{ \mathcal{D}_{i}\right\} _{i=1}^{N}$.
In conventional Bayesian inference, the posterior of a $K$-dimensional
unknown parameter vector $\boldsymbol{\theta}$ combines a likelihood
$p\left(\mathcal{D}|\boldsymbol{\theta}\right)$ with a prior $p\left(\boldsymbol{\theta}\right)$,
\[
\pi\left(\boldsymbol{\theta}\right)\propto p\left(\mathcal{D}|\boldsymbol{\theta}\right)p\left(\boldsymbol{\theta}\right),
\]
where $\propto$ denotes a proportional relationship. In contrast,
generalized Bayesian inference introduces a generalized posterior
by incorporating a learning rate $\eta\left(>0\right)$ (also termed
a scaling parameter) alongside the likelihood, as follows:
\[
\pi_{\eta}^{*}\left(\boldsymbol{\theta}\right)\propto p\left(\mathcal{D}|\boldsymbol{\theta}\right)^{\eta}p\left(\boldsymbol{\theta}\right).
\]
By setting $\eta<1$, the posterior spread increases, rendering the
inference robust against model misspecifications \cite{Walker2001,Gruenwald2012,Miller2019,Miller2021}.
The second category of robust Bayesian methods involves Gibbs posterior
inference \cite{Zhang2006a,Zhang2006b,Jiang2008,Bissiri2016,Atchade2017,Syring2023}.
This method formulates an inferential problem using a generic loss
function $r_{i}\left(\boldsymbol{\theta};\mathcal{D}_{i}\right)$
rather than a probabilistic model. Under this approach, the posterior
is expressed as
\[
\pi_{\eta}^{*}\left(\boldsymbol{\theta}\right)\propto\exp\left\{ -Nr\left(\boldsymbol{\theta};\mathcal{D}\right)\right\} ^{\eta}p\left(\boldsymbol{\theta}\right),
\]
where $r\left(\boldsymbol{\theta};\mathcal{D}\right)$ is an empirical
risk function defined as 
\[
r\left(\boldsymbol{\theta};\mathcal{D}\right)=\frac{1}{N}\sum_{i=1}^{N}r_{i}\left(\boldsymbol{\theta};\mathcal{D}_{i}\right).
\]
The algorithm proposed in the study can be applied to both approaches.
In what follows, $q\left(\boldsymbol{\theta};\mathcal{D}\right)$
denotes either a likelihood $p\left(\mathcal{D}|\boldsymbol{\theta}\right)$
or pseudolikelihood $\exp\left\{ -Nr\left(\boldsymbol{\theta};\mathcal{D}\right)\right\} $,
with 
\begin{equation}
\pi_{\eta}^{*}\left(\boldsymbol{\theta}\right)\propto q\left(\boldsymbol{\theta};\mathcal{D}\right)^{\eta}p\left(\boldsymbol{\theta}\right).\label{eq: pseudo-posterior}
\end{equation}

There are various methods for selecting $\eta$ \cite{Gruenwald2017,Holmes2017,Lyddon2019,Syring2019},
each with distinct emphases. For a comparative analysis, refer to
\cite{Wu2023}. We aim to enhance the generalized posterior calibration
(GPC) algorithm \cite{Syring2019}. Within the GPC framework, the
coverage probability for a particular $\eta$ value is assessed using
bootstrap samples, and $\eta$ is chosen numerically to achieve the
specified frequentist coverage probability. However, a significant
drawback of the GPC method is its computational burden, as it necessiates
repeated execution of a posterior simulator on bootstrap samples until
convergence.

This paper proposes an improvement in the GPC that requires fewer
of iterations. This algorithm is inspired by sequential Monte Carlo
(SMC) samplers \cite{DelMoral2006}, exploiting the similarity between
the learning rate in the GPC and an inverse temperature in SMC sampling.
The target distribution with the current value of $\eta$ is approximated
using weighted particles, as in SMC sampling. The particle approximation
of the target distribution with a new value of $\eta$ is obtained
without posterior simulation as in the reweighting step in SMC sampling.
At least in theory, a desirable value of $\eta$ is reached in a single
step, although few iterations are required in practice due to the
particle degeneracy problem. 

Two recently proposed approaches are relevant to this paper; however,
they are limited in the dimension of the problem they can handle.
First, \cite{Tanakaforthcoming} proposed an algorithm that replaces
an MCMC sampler with an SMC sampler. The algorithm proposed in this
paper is different, in that it uses an MCMC sampler for posterior
simulation and improves on the optimization procedure. Whereas SMC
samplers are only applicable to relatively low-dimensional cases (dimension
less than 10); however, some MCMC samplers are feasible for high-dimensional
cases. Second, \cite{Frazier2024} proposed an approach that delivers
a well-calibrated generalized posterior without a learning rate parameter,
which is theoretically different from the posterior considered in
GPC. The construction of quasi-posteriors \cite{Frazier2024} is similar
to that of the quasi-posterior based on generalized method of moments
criterion (e.g., \cite{Chernozhukov2003}). Conducting Monte Carlo
simulations on this type of quasi-posterior is known to be very difficult
\cite{Yin2011} requiring that the posterior distribution be evaluated
each time and the inverse and determinant of a matrix of the same
dimension as the number of unknown parameters be computed. Therefore,
unlike the approach proposed in this paper, efficient posterior simulators
such as Gibbs samplers are not available for their framework.

The subsequent sections are organized as follows. Section 2 describes
the GPC algorithm and SMC samplers as background. Section 3 presents
the newly proposed algorithm. Its application to synthetic and real
data is illustrated in Section 4. Finally, Section 5 provides concluding
remarks.

\section{Background}

\subsection{Generalized posterior calibration}

This section describes the generalized posterior calibration (GPC)
\cite{Syring2019} (Figure 1). $\mathcal{C}_{\alpha}^{\eta}\left(\mathcal{D}\right)$
denotes the generalized posterior $100\left(1-\alpha\right)\%$ credible
set for $\boldsymbol{\theta}$ with $\eta$. The coverage probability
is expressed as
\[
c_{\alpha}\left(\eta|\mathbb{P}\right)=\mathbb{P}\left\{ \boldsymbol{\theta}^{\dagger}\left(\eta\right)\in\mathcal{C}_{\alpha}^{\eta}\left(\mathcal{D}\right)\right\} ,
\]
where $\boldsymbol{\theta}^{\dagger}\left(\eta\right)$ denotes the
Kullback--Leibler minimizer obtained with $\eta$. Given that the
true data distribution $\mathbb{P}$ is unknown, we substitute it
with the empirical distribution $\mathbb{P}_{N}$, 
\[
c_{\alpha}\left(\eta|\mathbb{P}_{N}\right)=\mathbb{P}_{N}\left\{ \hat{\boldsymbol{\theta}}\left(\eta\right)\in\mathcal{C}_{\alpha}^{\eta}\left(\mathcal{D}\right)\right\} ,
\]
where $\hat{\boldsymbol{\theta}}\left(\eta\right)$ represents a point
estimate of $\boldsymbol{\theta}$ obtained with $\eta$. However,
direct evaluation of $c_{\alpha}\left(\eta|\mathbb{P}_{N}\right)$
is infeasible because it requires exhaustive enumeration of all $N^{N}$
possible with-replacement samples from $\mathcal{D}$. Consequently,
we approximate $\mathbb{P}_{N}$ using a bootstrap method. With $B$
bootstrap samples $\left\{ \breve{\mathcal{D}}^{\left[b\right]}\right\} _{b=1}^{B}$,
the coverage probability is estimated as 

\[
\hat{c}_{\alpha}\left(\eta|\mathbb{P}_{N}\right)=\frac{1}{B}\sum_{b=1}^{B}\mathbb{I}\left\{ \hat{\boldsymbol{\theta}}\left(\eta\right)\in\mathcal{C}_{\alpha}^{\eta}\left(\breve{\mathcal{D}}^{\left[b\right]}\right)\right\} ,
\]
where $\mathbb{I}\left\{ \cdot\right\} $ denotes the indicator function.
The learning rate $\eta$ is selected by solving $\hat{c}_{\alpha}\left(\eta|\mathbb{P}_{N}\right)=1-\alpha$
via a stochastic approximation \cite{Robbins1951}. At the $s$th
iteration, a single step of stochastic approximation recursion is
applied:

\begin{equation}
\eta_{s+1}\leftarrow\eta_{s}+\varsigma_{l}\left[\hat{c}_{\alpha}\left(\eta_{s}|\mathbb{P}_{N}\right)-\left(1-\alpha\right)\right],\label{eq: stochastic approximation}
\end{equation}
where $\left\{ \varsigma_{l}\right\} $ is a nonincreasing sequence
such that $\sum_{l}\varsigma_{l}=\infty$ and $\sum_{l}\varsigma_{l}^{2}<\infty$.
While \cite{Syring2019} specified $\varsigma_{s}=s^{-0.51}$, we
adopt a variant of Keston's \cite{Kesten1958} rule (Figure 2), $\varsigma_{s}\left(l\right)=l^{-0.51}$,
where an auxiliary variable $l$ increases by one only when there
is a directional change in the trajectory of $\eta_{s}$ and $\hat{c}_{\alpha}\left(\eta_{s}|\mathbb{P}_{N}\right)<1$.
This adaptation significantly reduces the convergence time. 

\begin{figure}

\caption{Generalized posterior calibration}

\begin{raggedright}
\medskip{}
\par\end{raggedright}
\begin{raggedright}
\texttt{input:}observed dataset $\mathcal{D}$, target kernel $\pi_{\eta}^{*}\left(\cdot\right)$,
initial guess $\eta_{1}$, target credibility level $\alpha$, termination
threshold $\epsilon$.
\par\end{raggedright}
\begin{raggedright}
\medskip{}
\par\end{raggedright}
\begin{raggedright}
Generate $B$ bootstrap samples $\left\{ \breve{\mathcal{D}}^{\left[b\right]}\right\} _{b=1}^{B}$
from $\mathcal{D}$.
\par\end{raggedright}
\begin{raggedright}
Set $s\leftarrow1$ and $l\leftarrow1$.
\par\end{raggedright}
\begin{raggedright}
\texttt{while converge}
\par\end{raggedright}
\begin{raggedright}
$\quad$Compute $\hat{\boldsymbol{\theta}}$ with $\eta_{s}$.
\par\end{raggedright}
\begin{raggedright}
$\quad$\texttt{for} $b=1,...,B$\texttt{:}
\par\end{raggedright}
\begin{raggedright}
$\quad$$\quad$Simulate $M$ posterior draws for $\breve{\mathcal{D}}^{\left[b\right]}$
using an MCMC
\par\end{raggedright}
\begin{raggedright}
$\quad$$\quad$$\quad$sampler with $\eta_{s}$.
\par\end{raggedright}
\begin{raggedright}
$\quad$$\quad$Compute the credible set $\mathcal{C}_{\alpha}^{\eta_{s}}\left(\breve{\mathcal{D}}^{\left[b\right]}\right)$.
\par\end{raggedright}
\begin{raggedright}
$\quad$\texttt{end for}
\par\end{raggedright}
\begin{raggedright}
$\quad$Compute the coverage probability $\hat{c}_{\alpha}\left(\eta_{s}|\mathbb{P}_{N}\right)$.
\par\end{raggedright}
\begin{raggedright}
$\quad$\texttt{if} $\left|\hat{c}_{\alpha}\left(\eta_{s}|\mathbb{P}_{N}\right)-\left(1-\alpha\right)\right|<\epsilon$\texttt{:}
\par\end{raggedright}
\begin{raggedright}
$\quad$$\quad$Set $\hat{\eta}\leftarrow\eta_{s}$.
\par\end{raggedright}
\begin{raggedright}
$\quad$$\quad$\texttt{break}
\par\end{raggedright}
\begin{raggedright}
$\quad$\texttt{else}
\par\end{raggedright}
\begin{raggedright}
$\quad$$\quad$Set a new learning rate $\eta_{s+1}$ according to
\par\end{raggedright}
\begin{raggedright}
$\quad$$\quad$$\quad$Figure 2 (GPC-SA) or Figure 3 (GPC-WP).
\par\end{raggedright}
\begin{raggedright}
$\quad$$\quad$Set $s\leftarrow s+1$.
\par\end{raggedright}
\begin{raggedright}
$\quad$\texttt{end if}
\par\end{raggedright}
\begin{raggedright}
\texttt{end while}
\par\end{raggedright}
\raggedright{}\texttt{return:} $\hat{\eta}$
\end{figure}

\begin{figure}

\caption{A single step of stochastic approximation recursion with Keston's
rule}

\medskip{}

\texttt{input:} last three values $\left\{ \eta_{i},\eta_{i-1},\eta_{i-2}\right\} $,
current value $l$, empirical coverage probability $\hat{c}_{\alpha}\left(\eta_{i}|\mathbb{P}_{N}\right)$,
credibility level $\alpha$

\medskip{}

\texttt{if} $\left(\eta_{i-1}-\eta_{i-2}\right)\left(\eta_{i}-\eta_{i-1}\right)<0$\texttt{
and $i>2$:}

$\quad$$l\leftarrow l+1$.

\texttt{end if}

Set a new learning rate:

\[
\eta_{i+1}\leftarrow\eta_{i}+l^{-0.51}\left[\hat{c}_{\alpha}\left(\eta_{i}|\mathbb{P}_{N}\right)-\left(1-\alpha\right)\right].
\]

\texttt{return:} $\eta_{i+1}$, $l$

\medskip{}

Note: $l$ is initialized to one at the beginning of the iterations.
\end{figure}

\subsection{Sequential Monte Carlo sampler}

The following describles sequential Monte Carlo (SMC) samplers \cite{DelMoral2006}.
These algorithms repeatedly utilize importance sampling to generate
$M$ weighted particles $\left\{ w^{\left[m\right]},\boldsymbol{\theta}^{\left[m\right]}\right\} _{m=1}^{M}$
that approximate a sequence of synthetic intermediate distributions
$\left\{ \pi_{t}\right\} _{t=0}^{T}$, where $\boldsymbol{\theta}^{\left[m\right]}$
denotes the $m$th particle and $w^{\left[m\right]}$ denotes the
corresponding weight. Initially, the distribution $\pi_{0}$ corresponds
to the prior, $\pi_{0}\left(\boldsymbol{\theta}\right)=p\left(\boldsymbol{\theta}\right)$,
while the terminal distribution $\pi_{T}$ is the target distribution,
i.e., the posterior, $\pi_{T}\left(\boldsymbol{\theta}\right)=p\left(\mathcal{D}|\boldsymbol{\theta}\right)p\left(\boldsymbol{\theta}\right)$.
Each intermediate distribution is defined as a likelihood-tempered
posterior:
\begin{equation}
\pi_{t}\left(\boldsymbol{\theta}\right)\propto p\left(\mathcal{D}|\boldsymbol{\theta}\right)^{\phi_{t}}p\left(\boldsymbol{\theta}\right),\label{eq: likelihood-tempered posterior}
\end{equation}
where $\left\{ \phi_{t}\right\} _{t=1}^{T}$ is an increasing sequence
with $\phi_{0}=0$ and $\phi_{T}=1$. The parameter $\phi_{t}$ can
be interpreted as the inverse temperature. We denote a sequence of
auxiliary distributions as $\left\{ \tilde{\pi}_{t}\right\} _{t=1}^{T}$
with 
\[
\tilde{\pi}_{t}\left(\boldsymbol{\theta}_{1:t}\right)=\pi_{t}\left(\boldsymbol{\theta}_{t}\right)\prod_{s=1}^{t-1}\mathcal{L}_{s}\left(\boldsymbol{\theta}_{s+1},\boldsymbol{\theta}_{s}\right),
\]
where $\mathcal{L}_{t}\left(\cdot,\cdot\right)$ denotes a Markov
kernel, also known as a backward kernel, which transitions backward
from $\boldsymbol{\theta}_{t+1}$ to $\boldsymbol{\theta}_{t}$. Weighted
particles $\left\{ w_{t}^{\left[m\right]},\boldsymbol{\theta}_{t}^{\left[m\right]}\right\} _{m=1}^{M}$
are used to approximate $\tilde{\pi}_{t}$. These particles are moved
via a Markov kernel $\mathcal{K}_{t}\left(\cdot,\cdot\right)$, also
termed a forward kernel. Let $\gamma_{t}\left(\boldsymbol{\theta}\right)$
denote the unnormalized posterior density with $\phi_{t}$,
\[
\gamma_{t}\left(\boldsymbol{\theta}\right)=p\left(\mathcal{D}|\boldsymbol{\theta}\right)^{\phi_{t}}p\left(\boldsymbol{\theta}\right).
\]
The sampling distribution of $\boldsymbol{\theta}_{1:t}^{\left[m\right]}=\left\{ \boldsymbol{\theta}_{1}^{\left[m\right]},...,\boldsymbol{\theta}_{t}^{\left[m\right]}\right\} $
is represented as:
\[
\zeta_{t}\left(\boldsymbol{\theta}_{1:t}^{\left[m\right]}\right)=\zeta_{1}\left(\boldsymbol{\theta}_{1}^{\left[m\right]}\right)\prod_{s^{\prime}=2}^{t}\mathcal{K}_{s^{\prime}}\left(\boldsymbol{\theta}_{s^{\prime}-1}^{\left[m\right]},\boldsymbol{\theta}_{s^{\prime}}^{\left[m\right]}\right).
\]
Importance sampling is applied to correct the discrepancy between
$\tilde{\pi}_{t}$ and $\zeta_{t}$. The unnormalized weights are
expressed as follows:
\[
W_{t}^{\left[m\right]}\propto\frac{\tilde{\pi}_{t}\left(\boldsymbol{\theta}_{1:t}^{\left[m\right]}\right)}{\zeta_{t}\left(\boldsymbol{\theta}_{1:t}^{\left[m\right]}\right)}=\frac{\tilde{\pi}_{t}\left(\boldsymbol{\theta}_{t}^{\left[m\right]}\right)\prod_{s=1}^{t-1}\mathcal{L}_{s}\left(\boldsymbol{\theta}_{s+1}^{\left[m\right]},\boldsymbol{\theta}_{s}^{\left[m\right]}\right)}{\zeta_{1}\left(\boldsymbol{\theta}_{1}^{\left[m\right]}\right)\prod_{s^{\prime}=2}^{t}\mathcal{K}_{s^{\prime}}\left(\boldsymbol{\theta}_{s^{\prime}-1}^{\left[m\right]},\boldsymbol{\theta}_{s^{\prime}}^{\left[m\right]}\right)}.
\]
Let $\widetilde{W}_{t}^{\left[m\right]}$ represent the unnormalized
incremental weight, defined as:
\[
\widetilde{W}_{t}^{\left[m\right]}=\frac{\gamma_{t}\left(\boldsymbol{\theta}_{t}^{\left[m\right]}\right)\mathcal{L}_{t-1}\left(\boldsymbol{\theta}_{t}^{\left[m\right]},\boldsymbol{\theta}_{t-1}^{\left[m\right]}\right)}{\gamma_{t-1}\left(\boldsymbol{\theta}_{t-1}^{\left[m\right]}\right)\mathcal{K}_{t}\left(\boldsymbol{\theta}_{t-1}^{\left[m\right]},\boldsymbol{\theta}_{t}^{\left[m\right]}\right)}.
\]
Then, we have $W_{t}^{\left[m\right]}\propto\widetilde{W}_{t}^{\left[m\right]}W_{t-1}^{\left[m\right]}$.

When the forward kernel $\mathcal{K}_{t}$ is chosen as $\pi_{t}$-invariant,
e.g., an MCMC kernel, this selection is deemed optimal because it
minimizes the variance of the weights \cite{DelMoral2006}. The backward
kernel is formulated as:
\[
\mathcal{L}_{t-1}\left(\boldsymbol{\theta}_{t}^{\left[m\right]},\boldsymbol{\theta}_{t-1}^{\left[m\right]}\right)=\frac{\pi_{t}\left(\boldsymbol{\theta}_{t-1}^{\left[m\right]}\right)\mathcal{K}_{t}\left(\boldsymbol{\theta}_{t-1}^{\left[m\right]},\boldsymbol{\theta}_{t}^{\left[m\right]}\right)}{\pi_{t}\left(\boldsymbol{\theta}_{t}^{\left[m\right]}\right)}.
\]
With this specification, the unnormalized incremental weights simplify
to:
\[
\widetilde{W}_{t}^{\left[m\right]}=\frac{\gamma_{t}\left(\boldsymbol{\theta}_{t-1}^{\left[m\right]}\right)}{\gamma_{t-1}\left(\boldsymbol{\theta}_{t-1}^{\left[m\right]}\right)}=p\left(\mathcal{D}|\boldsymbol{\theta}_{t-1}^{\left[m\right]}\right)^{\phi_{t}-\phi_{t-1}}.
\]
Thus, the unnormalized weights are updated as follows:
\[
W_{t}^{\left[m\right]}=w_{t-1}^{\left[m\right]}p\left(\mathcal{D};\boldsymbol{\theta}_{t-1}^{\left[m\right]}\right){}^{\phi_{t}-\phi_{t-1}}.
\]
The weights are normalized as: 
\[
w_{t}^{\left[m\right]}=W_{t}^{\left[m\right]}\left(\sum_{m^{\prime}=1}^{M}W_{t}^{\left[m^{\prime}\right]}\right)^{-1}.
\]
Note that this reweighting step does not involve a re-evaluation of
the likelihood $p\left(\mathcal{D}|\boldsymbol{\theta}_{t-1}^{\left[m\right]}\right)$.

As the difference in the consecutive temperatures increases, the variance
of the weights is likely to increase and only a portion of the weights
will be prominent, leading to the degeneration of the particle system.
To address this problem, the quality of the particle approximation
is monitored based on the effective sample size (ESS) \cite{Kong1994},
which captures the variance of the weights,
\[
ESS_{t}=\frac{1}{\sum_{m=1}^{M}\left(w_{t}^{\left[m\right]}\right)^{2}}=\frac{\left(\sum_{m=1}^{M}W_{t-1}^{\left[m\right]}\widetilde{W}_{t}^{\left[m\right]}\right)^{2}}{\sum_{m=1}^{M}\left(W_{t-1}^{\left[m\right]}\widetilde{W}_{t}^{\left[m\right]}\right)^{2}}.
\]
When the ESS is below a prespecified threshold $\overline{ESS}$,
the weighted particles are resampled \cite{Li2015}.

\section{Proposed Algorithm}

We propose a novel approach to find the best choice for the next learning
rate $\eta_{s+1}$, which is called weighted particle-based optimization.
The proposed approach is inspired by SMC samplers \cite{DelMoral2006,Dai2022}:
the learning rate $\eta$ in the generalized/Gibbs posterior (\ref{eq: pseudo-posterior})
and the inverse temperature in the likelihood-tempered posterior in
SMC sampling (\ref{eq: likelihood-tempered posterior}) play similar
roles in tempering the likelihood. 

At the $s$th iteration, we generate pseudoposterior draws using an
MCMC sampler with $\eta_{s}$. The posterior distribution with $\eta_{s}$
is approximated using a system of weighted particles, $\left\{ w_{s}^{\left[b,m\right]},\boldsymbol{\theta}_{s}^{\left[b,m\right]}\right\} _{m=1}^{M}$,
where $\sum_{m=1}^{M}w_{s}^{\left[b,m\right]}=1$ and each weight
$w_{s}^{\left[b,m\right]}$ is proportional to the posterior densities
evaluated at the corresponding posterior draw $\boldsymbol{\theta}_{s}^{\left[b,m\right]}$.
A new learning rate $\eta^{\prime}$ is chosen using stochastic approximation
recursion (Figure 2). As in the reweighting step in SMC sampling,
the weights for a new guess $\eta^{\prime}$ are computed as follows:
\[
\widetilde{W}_{s}^{\left[b,m\right]}=w_{s}^{\left[b,m\right]}q\left(\boldsymbol{\theta}_{s}^{\left[b,m\right]};\breve{\mathcal{D}}^{\left[b\right]}\right)^{\eta^{\prime}-\eta_{s}},
\]
\[
\widetilde{w_{s}}^{\left[b,m\right]}=\widetilde{W}_{s}^{\left[b,m\right]}\left(\sum_{m^{\prime}=1}^{M}\widetilde{W}_{s}^{\left[b,m^{\prime}\right]}\right)^{-1}.
\]
Therefore, we can evaluate the credible set for the new value $\eta^{\prime}$
without re-evaluating the likelihood. The credible interval for each
dimension is computed by sorting the particles and weights. Hereafter,
we refer to the GPC with stochastic approximation optimization as
the GPC-SA, while the GPC with weighted particle-based optimization
is the GPC-WP.

The advantage of the GPC-WP over the GPC-SA is its low computational
load. While the GPC-SA algorithm gradually approaches the optimal
value, the proposed algorithm reaches it directly; at least in theory,
no iteration is needed. However, as in SMC samplers, this algorithm
can suffer from particle degeneracy. To maintain the quality of particle
approximation, we terminate the stochastic approximation iteration
if the minimum ESS is below a prespecified threshold, \texttt{$minESS^{*}\left(\eta^{\prime};\eta_{s}\right)<\overline{ESS}$}
with\texttt{
\[
minESS^{*}\left(\eta^{\prime};\eta_{s}\right)=\min\left\{ ESS^{*\left[b\right]}\left(\eta^{\prime};\eta_{s}\right),\;b=1,...,B\right\} ,
\]
}
\[
ESS^{*\left[b\right]}\left(\eta^{\prime};\eta_{s}\right)=\frac{\left(\sum_{m=1}^{M}w_{s}^{\left[b,m\right]}q\left(\boldsymbol{\theta}_{s}^{\left[b,m\right]};\breve{\mathcal{D}}^{\left[b\right]}\right)^{\eta^{\prime}-\eta_{s}}\right)^{2}}{\sum_{m=1}^{M}\left(w_{s}^{\left[b,m\right]}p\left(\boldsymbol{\theta}_{s}^{\left[b,m\right]};\breve{\mathcal{D}}^{\left[b\right]}\right)^{\eta^{\prime}-\eta_{s}}\right)^{2}}.
\]
In this paper, we choose $\overline{ESS}=0.25M$. Figure 3 summarizes
the weighted particle-based optimization method, where parentheses
are added to the subscripts of the learning rate, as in $\eta_{\left(u\right)}^{\prime}$,
to distinguish it from the index related to the GPC iterations. $\iota$
denotes a flag for termination of the GPC algorithm: $\iota$ takes
the value 1 if a calibrated value is found and 0 otherwise. 

\begin{figure}

\caption{Weighted particle-based optimization}

\medskip{}

\texttt{input:} current learning rate $\eta_{s}$, current weighted
particles $\left\{ w_{s}^{\left[b,m\right]},\boldsymbol{\theta}_{s}^{\left[b,m\right]}\right\} _{m=1}^{M}$,
resampling threshold $\overline{ESS}$, termination threshold $\epsilon$.

Set $u\leftarrow1$, $l\leftarrow1$, and $\iota\leftarrow0$.

\texttt{while not converged:}

$\quad$Set a new learning rate $\eta_{\left(u\right)}^{\prime}$
using Algorithm 2.

$\quad$Compute the posterior estimate $\hat{\boldsymbol{\theta}}$
with $\eta_{\left(u\right)}^{\prime}$

$\quad$\texttt{for} $b=1,...,B$\texttt{:}

$\quad$$\quad$Compute the unnormalized weights under $\eta_{\left(u\right)}^{\prime}$:
\[
\widetilde{W}^{\left[b,m\right]}\leftarrow w_{s}^{\left[b,m\right]}q\left(\boldsymbol{\theta}_{s}^{\left[b,m\right]};\breve{\mathcal{D}}^{\left[b\right]}\right)^{\eta_{\left(u\right)}^{\prime}-\eta_{s}}
\]
.

$\quad$$\quad$Normalize the weights:
\[
\widetilde{w}^{\left[b,m\right]}\leftarrow\widetilde{W}^{\left[b,m\right]}\left(\sum_{m^{\prime}=1}^{M}\widetilde{W}^{\left[b,m^{\prime}\right]}\right)^{-1}.
\]

$\quad$$\quad$Compute the credible set $\mathcal{C}_{\alpha}^{\eta_{\left(u\right)}^{\prime}}\left(\breve{\mathcal{D}}^{\left[b\right]}\right)$.

$\quad$\texttt{end for}

$\quad$\texttt{if} $\left|\hat{c}_{\alpha}\left(\eta_{\left(u\right)}^{\prime}|\mathbb{P}_{N}\right)-\left(1-\alpha\right)\right|<\epsilon$\texttt{
and }

$\quad$$\quad$$\quad$\texttt{$minESS^{*}\left(\eta_{\left(u\right)}^{\prime};\eta_{s}\right)\geq\overline{ESS}$
then:}

$\quad$$\quad$A calibrated value has been reached; terminate GPC:

$\quad$$\quad$$\hat{\eta}\leftarrow\eta_{\left(u\right)}^{\prime}$
and $\iota\leftarrow1$. 

$\quad$\texttt{else if} \texttt{$minESS^{*}\left(\eta_{\left(u\right)}^{\prime};\eta_{s}\right)<\overline{ESS}$
then:}

$\quad$$\quad$Set the next value and continue GPC:

$\quad$$\quad$$\eta_{s+1}\leftarrow\eta_{\left(u\right)}^{\prime}$.

$\quad$$\quad$\texttt{break}

$\quad$\texttt{else:}

$\quad$$\quad$Set $u\leftarrow u+1$.

$\quad$\texttt{end if}

\texttt{end while}

\texttt{return:} $\hat{\eta}$ or $\eta_{s+1}$, as well as $\iota$

\end{figure}

\section{Application}

\subsection{Misspecified linear regression}

We considered a misspecified linear regression with synthetic data.
The model to be estimated was specified as a homoskedastic linear
regression: 
\[
y_{i}=\boldsymbol{\beta}^{\top}\boldsymbol{x}_{i}+\varepsilon_{i},\quad\varepsilon_{i}\sim\mathcal{N}\left(0,\sigma^{2}\right),
\]
where $y_{i}$ denotes a response variable; $\boldsymbol{x}_{i}=\left(1,x_{1,i},x_{2,i},x_{3,i}\right)^{\top}$
denotes a vector of covariates; $\boldsymbol{\beta}$ is the corresponding
coefficient vector; $\varepsilon_{i}$ denotes an error term with
variance $\sigma^{2}$; $\mathcal{N}\left(a,b^{2}\right)$ denotes
a normal distribution with mean $a$ and variance $b^{2}$. The true
data generation process, which was borrowed from Section 5.2 of \cite{Wu2023}
(Degree 2), had heteroskedastic errors depending on $x_{1,i}$:
\[
y_{i}=\boldsymbol{\beta}^{\top}\boldsymbol{x}_{i}+\varepsilon_{i},\quad\varepsilon_{i}\sim\mathcal{N}\left(0,\sigma_{i}^{2}\right),
\]
where $\sigma_{i}^{2}$ is 0.05 if $x_{1,i}<\xi_{0.05}$, 0.25 if
$\xi_{0.05}\leq x_{1,i}\leq\xi_{0.95}$, and 1 if $\xi_{0.95}<x_{1,i}$;
$\xi_{0.05}$ and $\xi_{0.95}$ denote the $5$th and 95th sample
percentiles of $\left\{ x_{1,1},,...,x_{1,N}\right\} $; and $\boldsymbol{\beta}=\left(\beta_{1},\beta_{2},\beta_{3},\beta_{4}\right)^{\top}=\left(1,1,2,-1\right)^{\top}$.
Three sample sizes $N\in\left\{ 100,200,400\right\} $ were examined.

Posterior simulations were conducted using a Gibbs sampler. The unknown
parameters $\boldsymbol{\theta}=\left(\boldsymbol{\beta}^{\top},\sigma^{2}\right)^{\top}$
were inferred using a normal prior, $\boldsymbol{\beta}\sim\mathcal{N}\left(\boldsymbol{0}_{4},\varsigma^{2}\boldsymbol{I}_{4}\right)$,
and an inverse gamma prior, $\sigma^{2}\sim\mathcal{IG}\left(\varrho_{1},\varrho_{2}\right)$,
where $\boldsymbol{0}_{A}$ denotes the $A$-dimensional vector of
zeros, $\boldsymbol{I}_{A}$ denotes the $A$-dimensional identity
matrix, $\varsigma^{2}$, $\varrho_{1}$, and $\varrho_{2}$ are hyperparameters;
$\mathcal{IG}\left(a,b\right)$ denotes an inverse gamma distribution
with shape parameter $a$ and rate parameter $b$. The hyperparameters
for $\sigma^{2}$ were set as $\left(\varrho_{1},\varrho_{2}\right)=\left(1,0.025\right)$.
While \cite{Wu2023} specified $\varsigma^{2}=\sigma^{2}$, $\beta_{i}\sim\mathcal{N}\left(0,\sigma^{2}\right)$,
our prior of $\boldsymbol{\beta}$ was independent of $\sigma^{2}$,
$\varsigma^{2}=100$. Consequently, the resulting posterior was more
general and computationally expensive because the inverse matrix had
to be computed repeatedly. 

We chose $\alpha=0.05$: the calibration target was the 95\% credible
set. Five hundred bootstrap samples were generated, so $B=500$. The
stopping criterion was $\epsilon=0.005$. All the simulations began
with $\eta=1$. The programs were executed on MATLAB (R2023b) on an
Ubuntu desktop (22.04.4 LTS) running on an AMD Ryzen Threadripper
3990X 2.9 GHz 64-core processor. The computations using different
bootstrap samples were parallelized.

We applied the two algorithms to 200 synthetic datasets. 3,000 posterior
draws were simulated and the last 2,000 draws were used for posterior
analysis. As shown in Table I, both algorithms effectively achieved
the target credibility level. Table II and Figure 4 compares the computational
costs of the algorithms. The panels in the left column report the
wall clock time in seconds. The median computation for the GPC-WP
was approximately 25-30\% faster than that of the GPC-SA. The improvement
in the computational efficiency was due to the smaller number of iterations,
as shown in the right two columns of Table III and the the panels
in the right column of Figure 4. While the GPC-SA needed approximately
nine iterations, the GPC-WP converged within two or three iterations
in most cases. The reason why the GPC-WP was not as fast as the GPC-SA
in terms of the number of iterations was that it took more time to
obtain the next learning rates. Nevertheless, this additional computational
cost was outweighed by the gain from reducing the number of iterations. 

\begin{table}
\caption{Coverage probability}

\medskip{}

\begin{centering}
\begin{tabular}{rrr}
\hline 
\multicolumn{1}{c}{$N$} & \multicolumn{2}{c}{Coverage probability (\%)}\tabularnewline
\cline{2-3} \cline{3-3} 
 & GPC-SA & GPC-WP\tabularnewline
\hline 
100 & 96.0 & 95.0\tabularnewline
200 & 96.0 & 95.5\tabularnewline
400 & 95.5 & 94.0\tabularnewline
\hline 
\end{tabular}
\par\end{centering}
\medskip{}

\centering{}%
\begin{minipage}[t]{0.9\columnwidth}%
Note: The coverage probability of the ground truth is evaluated based
on 200 synthetic datasets.%
\end{minipage}
\end{table}

\begin{table}

\caption{Comparison of the computational cost}

\medskip{}

\begin{centering}
\begin{tabular}{rcccc}
\hline 
\multicolumn{1}{c}{$N$} & \multicolumn{2}{c}{Computation time (sec.)} & \multicolumn{2}{r}{Number of iteration}\tabularnewline
\cline{2-5} \cline{3-5} \cline{4-5} \cline{5-5} 
 & GPC-SA & GPC-WP & GPC-SA & GPC-WP\tabularnewline
\hline 
100 & 12.3 & 8.8 & 9 & 3\tabularnewline
 & (9.7, 13.8) & (7.9, 10.0) & (7,10) & (2,3)\tabularnewline
200 & 12.4 & 9.2 & 9 & 3\tabularnewline
 & (11.0, 13.8) & (8.4, 10.3) & (8,10) & (2,3)\tabularnewline
400 & 12.6 & 9.6 & 9 & 3\tabularnewline
 & (11.2, 14.0) & (8.6, 10.4) & (8,10) & (2,3)\tabularnewline
\hline 
\end{tabular}
\par\end{centering}
\medskip{}

\centering{}%
\begin{minipage}[t]{0.9\columnwidth}%
Note: The median value is computed based on 200 synthetic datasets.
The numbers in parenthesis denote the interquartile range.%
\end{minipage}
\end{table}

\begin{figure*}
\caption{Comparison of the computational cost}

\medskip{}

\begin{centering}
\includegraphics{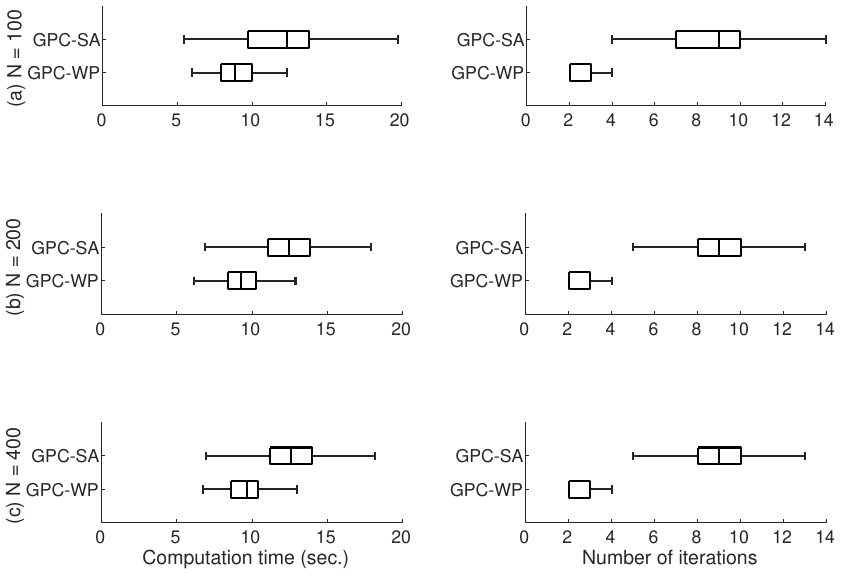}
\par\end{centering}
\medskip{}

\centering{}%
\begin{minipage}[t]{0.9\columnwidth}%
Note: The boxplot displays the distribution of computation time (in
seconds) for 200 synthetic datasets.%
\end{minipage}
\end{figure*}

\subsection{Support vector machine}

We applied the proposed algorithm to estimate a support vector machine
with real data, following Section 5 of \cite{Syring2019}. The outcome
is binary, $y_{i}\in\left\{ -1,1\right\} $, and the feature vector
is composed of an intercept and $K-1$ covariates, $\boldsymbol{x}_{i}=\left(1,x_{1,i},...,x_{K-1,i}\right)^{\top}$.
Estimating the support vector classifier is represented as the problem
of minimizing the following empirical risk function: 
\[
r\left(\boldsymbol{\theta}\right)=\frac{1}{N}\sum_{i=1}^{N}2\max\left(0,1-y_{i}\boldsymbol{\theta}^{\top}\boldsymbol{x}_{i}\right),
\]
where $\boldsymbol{\theta}\in\mathbb{R}^{K}$ is a vector of unknown
parameters. We assigned an independent Laplace-type prior to $\boldsymbol{\theta}$.
Then, the log pseudoposterior is represented as
\[
\pi_{\eta}^{*}\left(\boldsymbol{\theta}\right)\propto-\eta\sum_{i=1}^{N}2\max\left(0,1-y_{i}\boldsymbol{x}_{i}^{\top}\boldsymbol{\theta}\right)-\nu^{-1}\sum_{k=1}^{K}\left|\frac{\theta_{k}}{\sigma_{k}}\right|,
\]
where $\sigma_{k}$ denotes the standard deviation of the $k$th predictor
$x_{k,1},...,x_{k,N}$ with $\sigma_{1}=1$, and $\nu\left(>0\right)$
is a hyperparameter. We chose $\nu=10$. This model was inferred for
the South African Heart Disease dataset as in Section 4.4.2 of \cite{Hastie2009},
with $N=462$ and $K=8$. For the posterior simulation, we modified
a Gibbs sampler of \cite{Polson2011} by incorporating $\eta$.

The two algorithms were executed 50 times using different random seeds.
We generated total 11,000 draws, discarding the first 1,000 draws
as warmup. For every execution, both algorithms reached $\eta\approx0.09$,
which is in agreement with \cite{Syring2019}. Table III and Figure
5 report the computation time and number of iterations. The median
computation time for the GPC-SA was 201.7 seconds, while that for
the GPC-WP was 124.4 seconds (38.3\% faster). The median required
number of iterations for the GPC-SA was 7, while that for the GPC-WP
was 4. 

\begin{table}
\caption{Comparison of the computational cost}

\medskip{}

\begin{centering}
\begin{tabular}{cccc}
\hline 
\multicolumn{2}{c}{Computation time (sec.)} & \multicolumn{2}{r}{Number of iteration}\tabularnewline
\hline 
GPC-SA & GPC-WP & GPC-SA & GPC-WP\tabularnewline
\hline 
201.7 & 124.4 & 7 & 4\tabularnewline
(144.6, 231.4) & (93.2, 183.9) & (5, 8) & (3, 5)\tabularnewline
\hline 
\end{tabular}
\par\end{centering}
\medskip{}

\centering{}%
\begin{minipage}[t]{0.9\columnwidth}%
Note: The median value is computed based on 200 synthetic datasets.
The numbers in parenthesis denote the interquartile range.%
\end{minipage}
\end{table}

\begin{figure*}
\caption{Comparison of the computational cost}

\medskip{}

\begin{centering}
\includegraphics{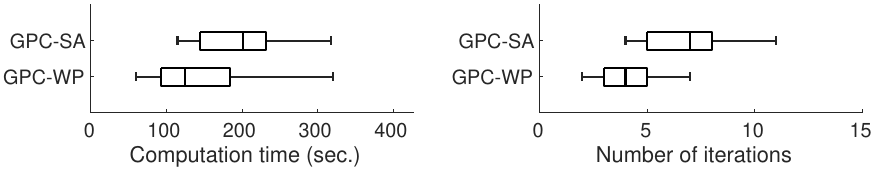}
\par\end{centering}
\medskip{}

\centering{}%
\begin{minipage}[t]{0.9\columnwidth}%
Note: The boxplot displays the distribution of computation time (in
seconds) for 50 runs with different random seeds.%
\end{minipage}
\end{figure*}

\section{Conclusion}

In this paper, we explored two robust Bayesian methods: generalized
Bayesian inference and Gibbs posterior inference. These methods offer
robustness against model misspecifications by introducing learning
rate parameters and formulating inferential problems using generic
loss functions. Central to these approaches is the calibration of
the posterior distribution spread through the selection of an appropriate
learning rate parameter. Building upon the GPC algorithm proposed
by \cite{Syring2019}, we introduced a more efficient version inspired
by SMC samplers. Our proposed algorithm evaluates the coverage probability
with a different learning rate without the need for repeated posterior
simulations, substantially reducing the computational costs while
obtaining the desirable values within a few iterations. Through synthetic
and real data applications, we demonstrated the efficacy of our proposed
algorithm. By providing a computationally efficient solution for learning
rate selection, our work contributes to advancing robust Bayesian
methods in statistical learning, facilitating their practical applicability
in complex modeling scenarios. 

\bibliographystyle{ieeetr}
\bibliography{reference}

\begin{thebibliography}{10}

\bibitem{Syring2019}
N.~Syring and R.~Martin, ``Calibrating general posterior credible regions,''
  {\em Biometrika}, vol.~106, no.~2, pp.~479--486, 2019.

\bibitem{Martin2022}
R.~Martin and N.~Syring, ``Direct {G}ibbs posterior inference on risk
  minimizers: Construction, concentration, and calibration,'' in {\em
  Advancements in Bayesian Methods and Implementation} (A.~S. {Srinivasa Rao},
  G.~A. Young, and C.~Rao, eds.), vol.~47 of {\em Handbook of Statistics},
  ch.~1, pp.~1--41, Elsevier, 2022.

\bibitem{Nottforthcoming}
D.~J. Nott, C.~Drovandi, and D.~T. Frazier, ``Bayesian inference for
  misspecified generative models,'' {\em Annual Review of Statistics and Its
  Application}, forthcoming.

\bibitem{Walker2001}
S.~Walker and N.~L. Hjort, ``On {B}ayesian consistency,'' {\em Journal of the
  Royal Statistical Society Series B: Statistical Methodology}, vol.~63, no.~4,
  pp.~811--821, 2001.

\bibitem{Gruenwald2012}
P.~Gr{\"u}nwald, ``The safe {B}ayesian,'' in {\em Algorithmic Learning Theory}
  (N.~H. Bshouty, G.~Stoltz, N.~Vayatis, and T.~Zeugmann, eds.), (Berlin,
  Heidelberg), pp.~169--183, Springer Berlin Heidelberg, 2012.

\bibitem{Miller2019}
J.~W. Miller and D.~B. Dunson, ``Robust {B}ayesian inference {via}
  coarsening,'' {\em Journal of the American Statistical Association},
  vol.~114, no.~527, pp.~1113--1125, 2019.

\bibitem{Miller2021}
J.~W. Miller, ``Asymptotic normality, concentration, and coverage of
  generalized posteriors,'' {\em Journal of Machine Learning Research},
  vol.~22, no.~168, pp.~1--53, 2021.

\bibitem{Zhang2006a}
T.~Zhang, ``Information-theoretic upper and lower bounds for statistical
  estimation,'' {\em {IEEE} Transactions on Information Theory}, vol.~52,
  no.~4, pp.~1307--1321, 2006.

\bibitem{Zhang2006b}
T.~Zhang, ``From $\varepsilon$-entropy to {KL}-entropy: Analysis of minimum
  information complexity density estimation,'' {\em Annals of Statistics},
  vol.~34, no.~5, pp.~2180--2210, 2006.

\bibitem{Jiang2008}
W.~Jiang and M.~A. Tanner, ``Gibbs posterior for variable selection in
  high-dimensional classification and data mining,'' {\em Annals of
  Statistics}, vol.~36, no.~5, pp.~2207--2231, 2008.

\bibitem{Bissiri2016}
P.~G. Bissiri, C.~C. Holmes, and S.~G. Walker, ``A general framework for
  updating belief distributions,'' {\em Journal of the Royal Statistical
  Society Series B: Statistical Methodology}, vol.~78, no.~5, pp.~1103--1130,
  2016.

\bibitem{Atchade2017}
Y.~A. Atchad{\'e}, ``On the contraction properties of some high-dimensional
  quasi-posterior distributions,'' {\em Annals of Statistics}, vol.~45, no.~5,
  pp.~2248--2273, 2017.

\bibitem{Syring2023}
N.~Syring and R.~Martin, ``Gibbs posterior concentration rates under
  sub-exponential type losses,'' {\em Bernoulli}, vol.~29, no.~2,
  pp.~1080--1108, 2023.

\bibitem{Gruenwald2017}
P.~Gr{\"u}nwald and T.~van Ommen, ``Inconsistency of {B}ayesian inference for
  misspecified linear models, and a proposal for repairing it,'' {\em Bayesian
  Analysis}, vol.~12, no.~4, pp.~1069--1103, 2017.

\bibitem{Holmes2017}
C.~C. Holmes and S.~G. Walker, ``Assigning a value to a power likelihood in a
  general {B}ayesian model,'' {\em Biometrika}, vol.~104, no.~2, pp.~497--503,
  2017.

\bibitem{Lyddon2019}
S.~P. Lyddon, C.~C. Holmes, and S.~G. Walker, ``General {B}ayesian updating and
  the loss-likelihood bootstrap,'' {\em Biometrika}, vol.~106, no.~2,
  pp.~465--478, 2019.

\bibitem{Wu2023}
P.-S. Wu and R.~Martin, ``A comparison of learning rate selection methods in
  generalized {B}ayesian inference,'' {\em Bayesian Analysis}, vol.~18, no.~1,
  pp.~105--132, 2023.

\bibitem{DelMoral2006}
P.~Del~Moral, A.~Doucet, and A.~Jasra, ``Sequential {M}onte {C}arlo samplers,''
  {\em Journal of the Royal Statistical Society Series B: Statistical
  Methodology}, vol.~68, no.~3, pp.~411--436, 2006.

\bibitem{Tanakaforthcoming}
M.~Tanaka, ``Generalized posterior calibration via sequential monte carlo
  sampler,'' in {\em Proceedings of the 2024 6th Asia Conference on Machine
  Learning and Computing}, forthcoming.

\bibitem{Frazier2024}
D.~T. Frazier, C.~Drovandi, and R.~Kohn, ``Calibrated generalized {B}ayesian
  inference,'' tech. rep., arXiv preprint, arXiv:2311.15485, 2024.

\bibitem{Chernozhukov2003}
V.~Chernozhukov and H.~Hong, ``An {MCMC} approach to classical estimation,''
  {\em Journal of Econometrics}, vol.~115, no.~2, pp.~293--346, 2003.

\bibitem{Yin2011}
G.~Yin, Y.~Ma, F.~Liang, and Y.~Yuan, ``Stochastic generalized method of
  moments,'' {\em Journal of Computational and Graphical Statistics}, vol.~20,
  no.~3, pp.~714--727, 2011.

\bibitem{Robbins1951}
H.~Robbins and S.~Monro, ``A stochastic approximation method,'' {\em Annals of
  Mathematical Statistics}, vol.~22, no.~3, pp.~400--407, 1951.

\bibitem{Kesten1958}
H.~Kesten, ``Accelerated stochastic approximation,'' {\em Annals of
  Mathematical Statistics}, vol.~29, no.~1, pp.~41--59, 1958.

\bibitem{Kong1994}
A.~Kong, J.~S. Liu, and W.~H. Wong, ``Sequential imputations and {B}ayesian
  missing data problems,'' {\em Journal of the American Statistical
  Association}, vol.~89, no.~425, pp.~278--288, 1994.

\bibitem{Li2015}
T.~Li, M.~Bolic, and P.~M. Djuric, ``Resampling methods for particle filtering:
  Classification, implementation, and strategies,'' {\em {IEEE} Signal
  Processing Magazine}, vol.~32, no.~3, pp.~70--86, 2015.

\bibitem{Dai2022}
C.~Dai, P.~E.~J. Jeremy~Heng, and N.~Whiteley, ``An invitation to sequential
  {M}onte {C}arlo samplers,'' {\em Journal of the American Statistical
  Association}, vol.~117, no.~539, pp.~1587--1600, 2022.

\bibitem{Hastie2009}
T.~Hastie, R.~Tibshirani, J.~H. Friedman, and J.~H. Friedman, {\em The Elements
  of Statistical Learning: Data Mining, Inference, and Prediction}.
\newblock Springer, 2~ed., 2009.

\bibitem{Polson2011}
N.~G. Polson and S.~L. Scott, ``Data augmentation for support vector
  machines,'' {\em Bayesian Analysis}, vol.~6, no.~1, pp.~43--48, 2011.

\end{thebibliography}

\end{document}